\documentclass{pasj00}
\draft

\begin{document}
\SetRunningHead{K. Yamada, H. Asada, M. Yamaguchi, and N. Gouda}
{Moment Approach to Astrometric Binary with Low SN} 
\Received{}
\Accepted{}

\title{Improving the Moment Approach for Astrometric Binaries:
Possible Application to Cygnus X-1}


\author{Kei \textsc{Yamada}\altaffilmark{1}, 
Masaki \textsc{Yamaguchi}\altaffilmark{2}, 
Hideki \textsc{Asada}\altaffilmark{1},
and 
Naoteru \textsc{Gouda}\altaffilmark{2}}
\altaffiltext{1}{Faculty of Science and Technology, Hirosaki University, 
Hirosaki, Aomori 036-8561}
\email{yamada@tap.st.hirosaki-u.ac.jp}
\altaffiltext{2}{National Astronomical Observatory of Japan, 
Mitaka, Tokyo, Japan 181-8588}


%


\maketitle

\begin{abstract}
A moment approach for orbit determinations of 
astrometric binaries from astrometric observations alone 
has been recently studied for a low signal-to-noise ratio
(Iwama et al. 2013, PASJ, {\bf 65}, 2). 
With avoiding a direct use of the time-consuming Kepler equation,
temporal information is taken into account to 
increase the accuracy of statistical moments.
As numerical tests, 100 realizations are done 
and the mean and the standard deviation are also evaluated.
For a semi-major axis, 
the difference between the mean of the recovered values 
and the true value decreases to less than a tenth 
in the case of $10000$ observed points.
Therefore, the present moment approach works better than the previous one
for the orbit determinations
when one has a number of the observed points.
The present approach is thus applicable to Cyg X-1.
\end{abstract}

\section{Introduction}
Space astrometry missions such as 
Gaia and JASMINE are expected to reach a few micro arcseconds
\citep{GAIA-1,GAIA-2,JASMINE}. 
Moreover, high-accuracy VLBI is also available. 

Orbit determinations for binaries have been considered for a long time.
For visual binaries, formulations for orbit determinations 
have been well developed since the nineteenth century 
\citep{Thiele,Binnendijk,Aitken,Danby,Roy}. 
At present, numerical methods are successfully used 
\citep{EX,CO,OC}. 
Furthermore, an analytic solution for an astrometric binary, 
where one object is unseen, 
has been found \citep{AAK,AAK2,Asada2008}. 
The solution requires that sufficiently accurate positions of a star 
(or a photocenter of the binary) are measured 
at more than four places
during an orbital cycle of the binary system.

A moment approach for a low signal-to-noise (SN) ratio 
is proposed by Iwama et al. (2013, hereafter the {\it Iwama+ approach}).
For a close binary system with a short orbital period,  
we have a relatively large uncertainty in the position measurements.
For instance, 
the orbital periods of Cyg X-1 and LS 5039 are 
nearly 6 days and 4 days, respectively,
which are extremely shorter than that of normal binary stars, 
say a few months and several years. 
Although temporal information is not incorporated in the Iwama+ approach,
this approach would be useful 
to obtain recovered values of orbital parameters, 
when observational errors are much smaller than a binary apparent size.
It would be convenient to use the recovered values as trial values of
the steepest descent method
for reaching the best-fit parameter values.

On the other hand, if observational errors are 
comparable to or larger than 
a binary apparent size,
the orbital parameters cannot be recovered well,
because the expected values of the statistical moments are 
quite different from the true values.
Hence, it is important to improve the Iwama+ approach
in order to treat such a case of extremely low SN ratio.
The main purpose of this paper 
is to improve the previous approach 
by using temporal information of observed points.
However, the use of the Kepler equation is still avoided
like the previous approach.

\section{Moment Formalism}

\begin{figure}
\includegraphics[width=84mm]{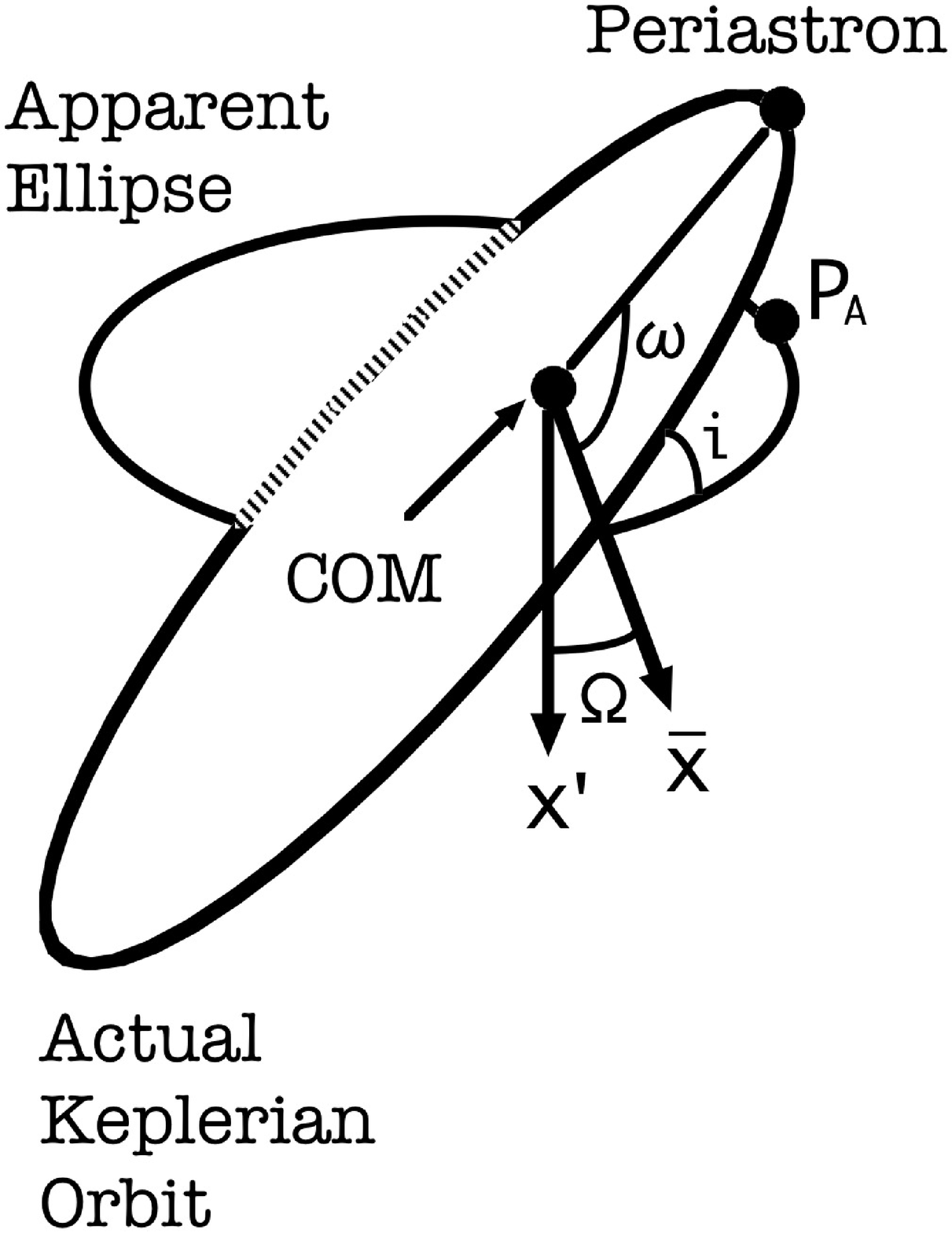}
\caption 
{Actual Keplerian orbit and apparent ellipse 
in three-dimensional space. 
We denote the inclination angle as $i$, 
the argument of periastron as $\omega$ and the longitude of 
ascending node as $\Omega$. 
These angles relate two coordinates $(x^{\prime}, y^{\prime})$ 
and $(\bar{x}, \bar{y})$, both of which choose the origin 
as the common center of mass. 
Here, the $x^{\prime}$ axis is taken to lie along the semi-major axis 
of the apparent ellipse, 
while the $\bar{x}$-axis is along 
the direction of the ascending node.
}
\label{config}
\end{figure}

We consider a Kepler orbit, 
whose semi-major axis, eccentricity, inclination angle, 
argument of periastron, and longitude of ascending node are 
$(a_K, e_K, i, \omega, \Omega)$ (see Fig. \ref{config}).
Here, we focus on a binary whose orbital period $P_K$ is known by 
other observations.
Angular positions projected onto the celestial sphere are expressed 
by using the Thiele-Innes elements \citep{Aitken, Binnendijk, Roy}.

Let us assume frequent observations of 
the angular position in the celestial sphere.
Namely, we consider a large number of observed points. 
For such a case, 
the statistical average expressed as a summation 
is taken as the temporal average in an integral form as 
\begin{equation}
< F > \equiv \frac{1}{T_{obs}} \int_{0}^{T_{obs}} F dt , 
\label{average1}
\end{equation}
where $< \quad >$ denotes the mean and 
$T_{obs}$ denotes the total time duration of the observations.  

In this paper, we focus on the periodic motion, 
so that the above expression becomes 
the integration over 
several orbital periods.
We thus obtain 
\begin{eqnarray}
< F > &=& \frac{1}{J P} \int_{t_0}^{t_0+J P} F dt  
\nonumber\\
&=& \frac{1}{P} \int_{t_0}^{t_0+P} F dt  
\nonumber\\
&=& 
\frac{1}{2\pi} \int_{0}^{2\pi} F (1 - e_K \cos u) du , 
\label{average2} 
\end{eqnarray}
where $J$ is an integer and we used the Kepler equation
\begin{equation}
t = t_0 + \frac{P_K}{2\pi} (u - e_K \sin u) ,
\label{Kepler}
\end{equation}
and $dt = P_K (1 - e_K \cos u) du /2 \pi $. 
Here, $u$ and $t_0$ denote the eccentric anomaly and 
the time of periastron passage, respectively.

Let us consider statistical moments. 
The second and the third moments of the projected position 
in $(x, y)$ coordinates are useful to determine orbital parameters.
They are defined as
\begin{eqnarray}
M_{xx} &\equiv& < (x-<x>)^2 > 
\nonumber\\
&=& \frac12 (\alpha^2+\beta^2) - \frac14 e_K^2 \alpha^2 , 
\label{Mxx}
\\
M_{yy} &\equiv& < (y-<y>)^2 > 
\nonumber\\
&=& \frac12 (\gamma^2+\delta^2) - \frac14 e_K^2 \gamma^2 , 
\label{Myy}
\\
M_{xy} &\equiv& < (x-<x>) (y-<y>) > 
\nonumber\\
&=& \frac12 (\alpha\gamma+\beta\delta) - \frac14 e_K^2 \alpha\gamma , 
\label{Mxy}
\\
M_{xxx} &\equiv& < (x-<x>)^3 > 
\nonumber\\
&=& \frac38 e_K \alpha  (\alpha^2+\beta^2) 
- \frac14 e_K^3 \alpha^3 , 
\label{Mxxx}
\\
M_{yyy} &\equiv& < (y-<y>)^3 > 
\nonumber\\
&=& \frac38 e_K \gamma (\gamma^2+\delta^2) 
- \frac14 e_K^3 \gamma^3 , 
\label{Myyy}
\\
M_{xxy} &\equiv& < (x-<x>)^2 (y-<y>) > 
\nonumber\\
&=& \frac18 e_K (3\alpha^2\gamma+\beta^2\gamma+2\alpha\beta\delta) 
- \frac14 e_K^3 \alpha^2\gamma , 
\label{Mxxy}
\\
M_{xyy} &\equiv& < (x-<x>) (y-<y>)^2 > 
\nonumber\\
&=& \frac18 e_K (3\alpha\gamma^2+\alpha\delta^2+2\beta\gamma\delta) 
- \frac14 e_K^3 \alpha\gamma^2 , 
\label{Mxyy}
\end{eqnarray}
where observational errors are assumed to vanish 
at the last equal in each equation,
and $\alpha$, $\beta$, $\gamma$, and $\delta$ are 
the Thiele-Innes type elements defined by \citep{IAY}
\begin{eqnarray}
\alpha &\equiv&
a_K (\cos\omega\cos\Omega - \sin\omega\sin\Omega \cos i) , 
\label{alpha}
\\
\beta &\equiv&
-b_K (\sin\omega\cos\Omega + \cos\omega\sin\Omega \cos i) , 
\label{beta}
\\
\gamma &\equiv&
a_K (\cos\omega\sin\Omega + \sin\omega\cos\Omega \cos i) , 
\label{gamma}
\\
\delta &\equiv&
-b_K (\sin\omega\sin\Omega - \cos\omega\cos\Omega \cos i) ,
\label{delta}
\end{eqnarray}
where $b_K = a_K \sqrt{1-e_K^2}$ is the semi-minor axis.
The moments $M_{xx}, \cdots, M_{xyy}$ are actually observables. 
For the moments calculation, temporal information 
of each observed position is smeared by averaging.
If positions of a star are measured 
with sufficiently small observation errors, 
one can recover the orbital parameters well 
by the Iwama+ approach \citep{IAY}.

\section{Improved Moment Approach}

\subsection{Observation errors}

In the above formalism, 
we assume that observed points are located on an apparent ellipse. 
However, position measurements are inevitably 
associated with observational errors. 
Therefore, it is very important to take into account observation noises.
In this paper, we add Gaussian errors into position measurements as 
$x \to x +\Delta x$ and $y \to y +\Delta y$, 
where $\Delta x$ and $\Delta y$ obey 
Gaussian distributions with a standard deviation $\sigma$. 
Then, the expected values of the moments are estimated as
\begin{eqnarray}
E(M_{xx}^{(O)}) &=& M_{xx}^{(T)} + \frac{N - 1}{N} \sigma^2 , 
\label{error-xx}\\
E(M_{yy}^{(O)}) &=& M_{yy}^{(T)} + \frac{N - 1}{N} \sigma^2 , 
\label{error-yy}\\
E(M_{xy}^{(O)}) &=& M_{xy}^{(T)} , \\
E(M_{xxx}^{(O)}) &=& M_{xxx}^{(T)} , \\
E(M_{yyy}^{(O)}) &=& M_{yyy}^{(T)} , \\
E(M_{xxy}^{(O)}) &=& M_{xxy}^{(T)} , \\
E(M_{xyy}^{(O)}) &=& M_{xyy}^{(T)} , 
\end{eqnarray}
where $N$ is the total number of observed points, and 
the upper indices $(O)$ and $(T)$ denote 
observables including observational errors 
and true values corresponding to Eqs. (\ref{Mxx}) - (\ref{Mxyy}), 
respectively.
Since $N$ is a large number, $(N-1)/N \simeq 1$.
Eqs. (\ref{error-xx}) and (\ref{error-yy}) suggest that 
orbital parameters are not recovered well
in the case that $\sigma^2$ is comparable to or larger than
$M_{xx}$ and $M_{yy}$,
even if $N$ approaches the infinity.
In this section, 
we improve the Iwama+ approach 
to obtain the moments with a higher accuracy
for such a large observational errors
by incorporating temporal information.

\subsection{Averaging operation}

By incorporating temporal information,
we average the coordinate values of observed points 
which are neighboring positions.
Let us assume that an orbital period of a binary $P_K$ is known
with high accuracy by another observation, 
such as observations of absorption lines
(e.g., \cite{BTLR} for Cyg X-1 and \cite{Sarty} for LS 5039).
If observational errors are so large, 
neighboring positions 
on the orbit can be considered 
as the same position 
within some errors.
In other words, 
one can identify an observed point at a time $t_1$
with another one at a time $t_2$ when 
\begin{equation}
\frac{\Delta t}{P_K} \ll \sigma, 
\end{equation}
in the units of $a_K = 1$, 
where
\begin{equation}
\Delta t = |t_1 - t_2| ~ (\mbox{mod } P_K) .
\end{equation}

Let us divide the apparent ellipse into small bins,
each of which corresponds to an equal short time interval,
e.g., $[t_0, t_0 + P_K/n_m]$, 
where $n_m$ is the number of the bins.
If the same star is observed at fixed intervals,
then, every bin has the equal number of observed points $n_a = N/n_m$
and one obtains more bins near the apastron than near periastron.
Namely, every bin will contain the same number of points 
if and only if the interval between the observations is not 
a multiple of the bin duration.
Note that we can use the data over several orbital periods, 
so that each bin may include observed points of different orbital periods.

\begin{figure}[h]
\includegraphics[width=84mm]{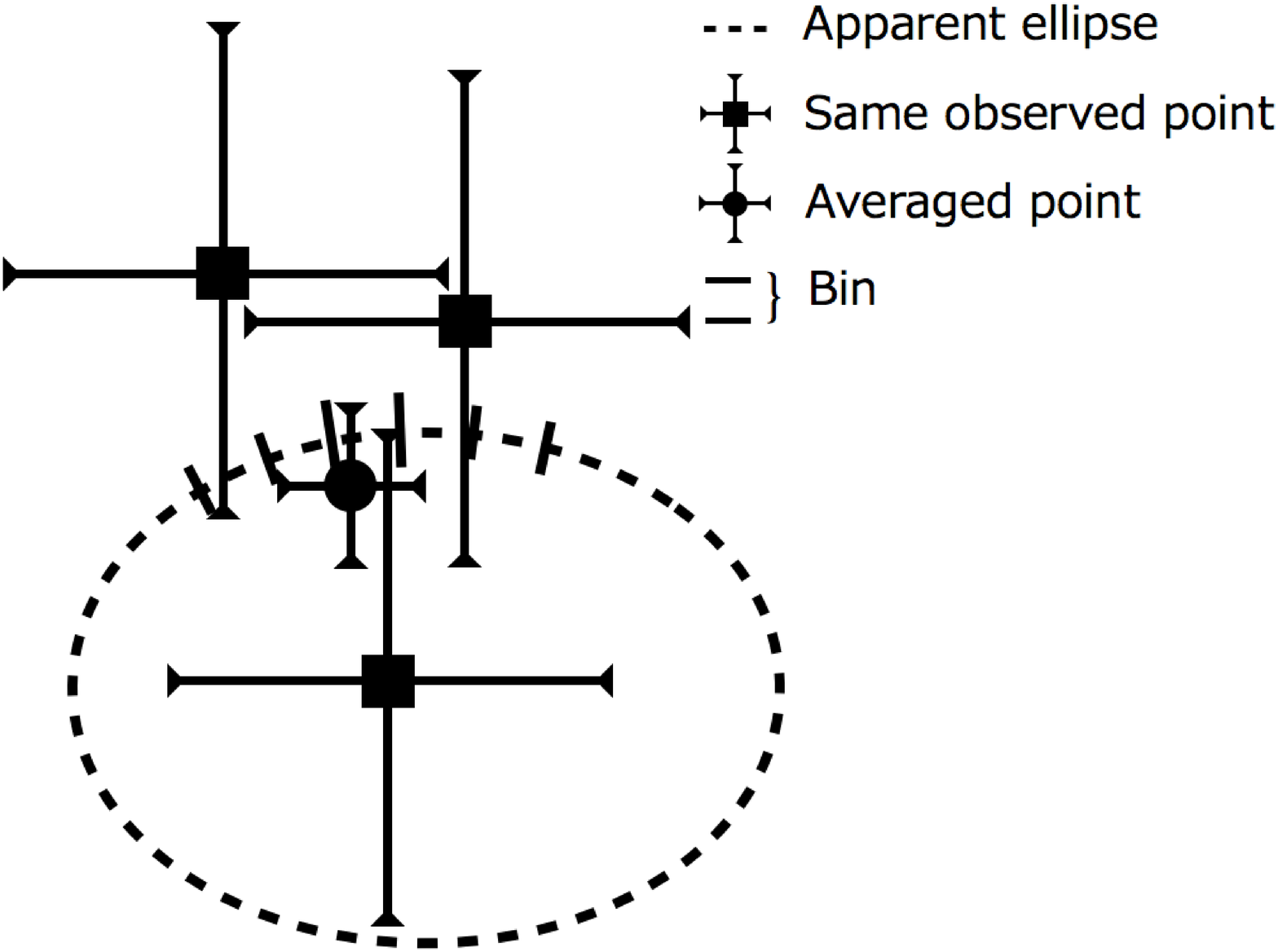}
\caption{The dashed ellipse is the apparent one, 
the square and disc with the error bars
are observed points and averaged one, 
respectively.
The apparent ellipse is divided 
by a small bin which has $n_a$ observed points.
The averaged point approaches the true value 
if $n_a$ is a large number.}
\label{fig-ave}
\end{figure}

In order to reduce statistical errors, 
we average the positions
of $n_a$ observed points for each bin 
and obtain $n_m$ averaged points
(see Fig. \ref{fig-ave}).
With this averaging operation, 
the expected values of the moments are given as
\begin{eqnarray}
E(\overline{M}_{xx}^{(O)}) &=& 
M_{xx}^{(T)} + \frac{N - 1}{N} \frac{\sigma^2}{n_a} , 
\label{average-xx}
\\
E(\overline{M}_{yy}^{(O)}) &=& 
M_{yy}^{(T)} + \frac{N - 1}{N} \frac{\sigma^2}{n_a} , 
\label{average-yy}
\end{eqnarray}
where the bar denotes the value obtained by
$n_m$ averaged points.
Hence, if $n_a$ is sufficiently large, 
the errors of $\overline{M}_{xx}^{(O)}$ and $\overline{M}_{yy}^{(O)}$ 
could be neglected safely.
Therefore, the Iwama+ approach is improved with regard to
the accuracy of the moments by the averaging operation.

\section{Results}

\subsection{Numerical test}

In Eqs. (\ref{average1}) and (\ref{average2}), 
we assume that one can integrate observed quantities.
In practice, however, observations are discrete, 
for which the integration should be replaced by a summation. 
The integration and the summation could agree 
in the limit that $n_m$ approaches the infinity. 
In addition,
it is necessary that $n_a$ is so large to reduce errors.
According to the numerical calculations, 
the present approach recovers
orbital parameters for $n_m = 100$ 
(see the discussion in \cite{IAY}
in the absence of the averaging operation, i.e. $n_a=1$ \& $n_m=N$).
Therefore, one can use $N / 100$ points 
for the averaging operation on each bin.

For the true parameters $(a_K, e_K, i, \omega, \Omega) 
= (1.0, 0.1, 30 \mathrm{~[deg.]}, 30 \mathrm{~[deg.]}, 30 \mathrm{~[deg.]})$, 
we consider two cases for $N=10000$.
Case 1: the observational error for each position measurement is 
equal to a binary size, namely, $\sigma = 1$ in the units of $a_K$. 
Case 2: $\sigma = 5$ in the units of $a_K$.
For each case, 
$\sigma/\sqrt{N}$ is fixed where we imagine an instrument, 
such as the Small-JASMINE.
For each parameter set, 100 realizations are done and the mean and 
the standard deviation are also evaluated.

\begin{figure}
\includegraphics[width=84mm]{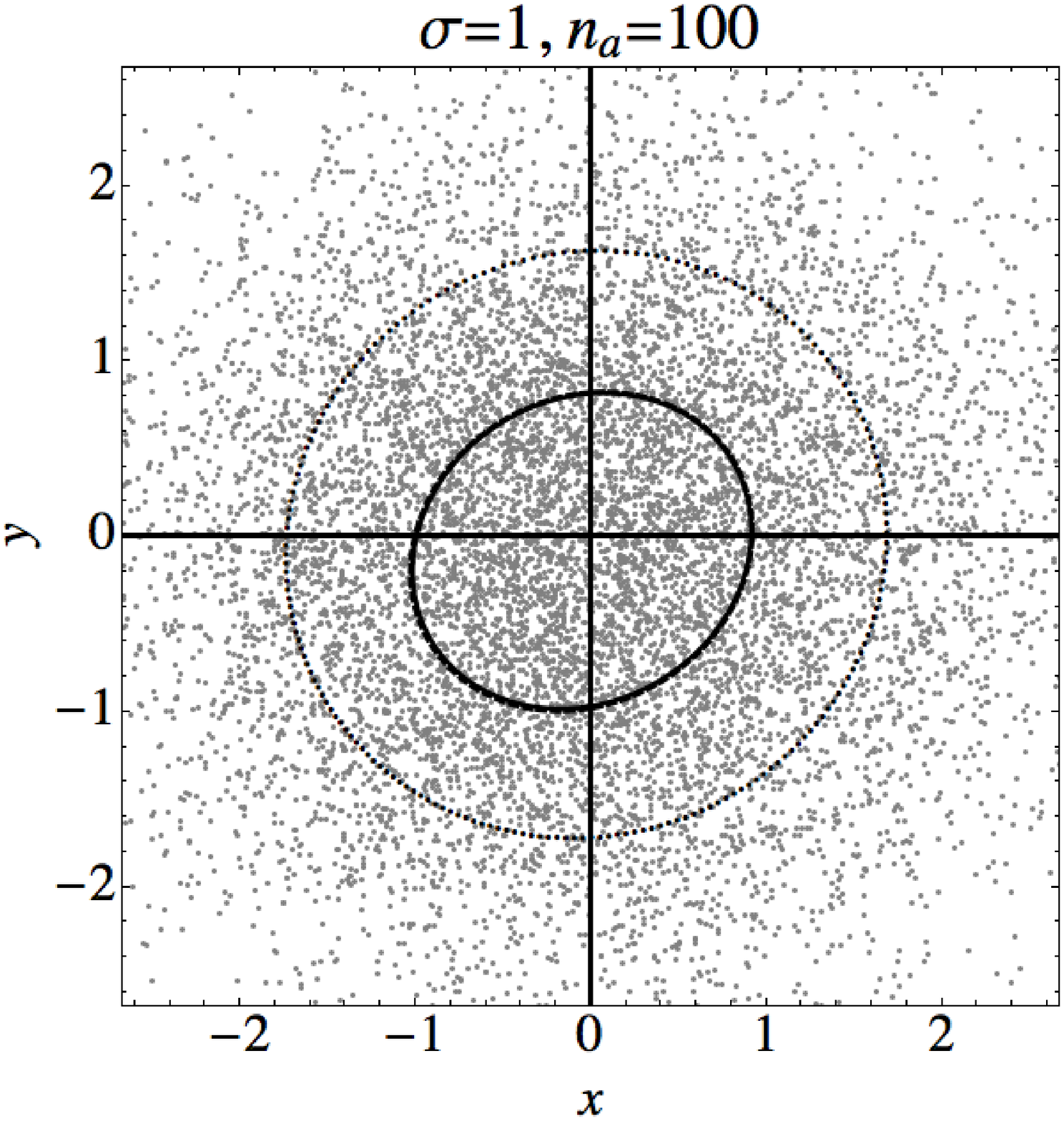}
\includegraphics[width=84mm]{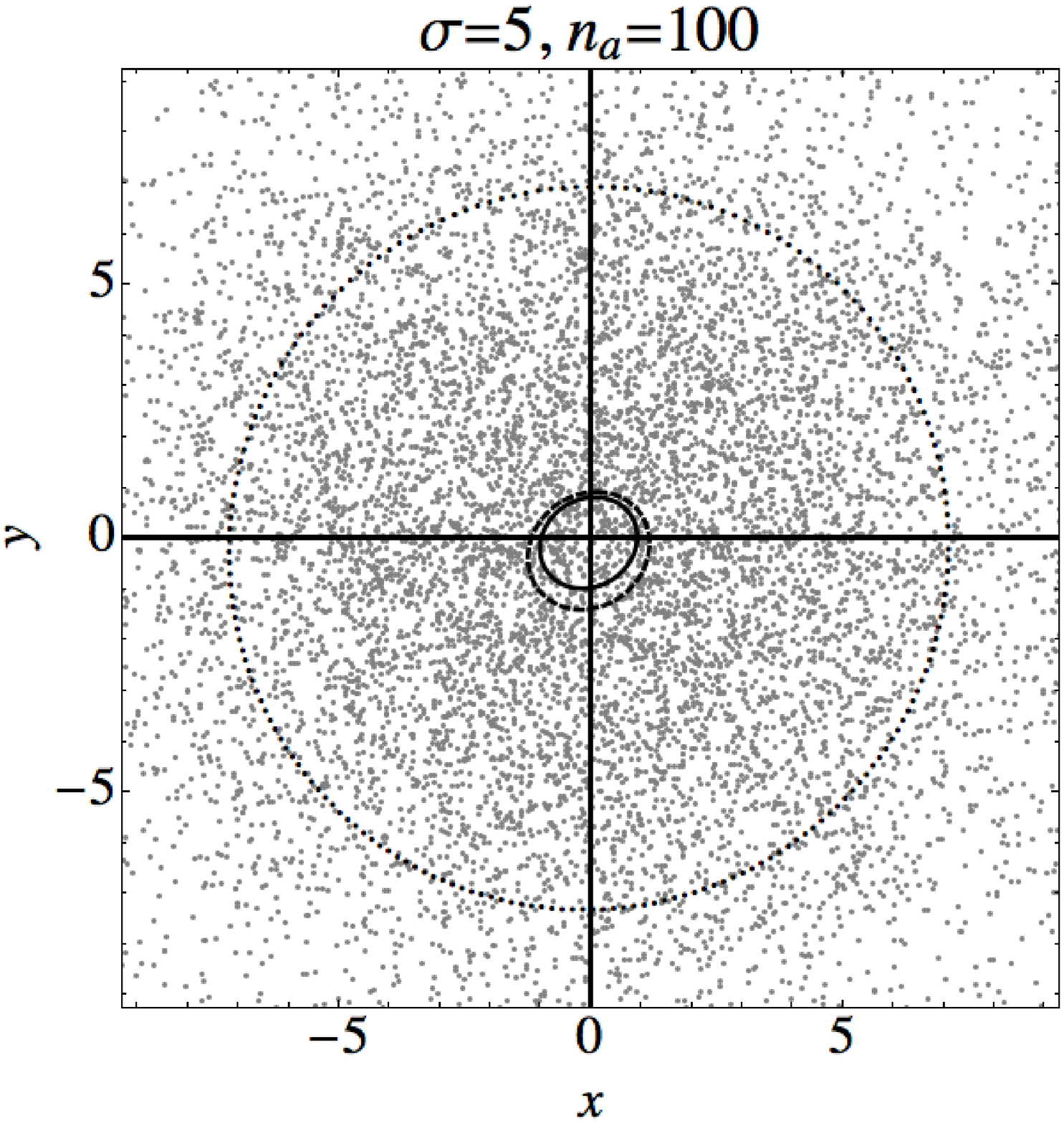}
\caption{
Numerical test: ten thousand observed points 
of the same source star on the $x-y$ plane. 
The parameters are $a_K = 1.0$, $e_K = 0.1$, 
$i = 30$ [deg.], $\omega = 30$ [deg.], $\Omega = 30$ [deg.], 
and $N=10000$ with $n_a =100$.
The gray points are the observed ones. 
The solid curve is the apparent ellipse for the true parameters.
The dotted and dashed curves denote orbits for the mean value 
of the recovered parameters by Iwama+ approach and the present one, 
respectively.
Left: $\sigma = 1$. Comparison between Iwama+ and the present approaches.
Here, the true orbit and the recovered orbit by the present approach
overlap each other.
Right: $\sigma = 5$. Comparison between Iwama+ and the present approaches. 
}
\label{comparing}
\end{figure}

Fig. \ref{comparing} shows the apparent orbits
for the mean values of the recovered parameters 
by Iwama+ approach and the present one.
The present approach can recover 
the orbital parameters better than the Iwama+ approach.
Especially, one can see that 
the true orbit and the recovered orbit by the present approach 
almost overlap each other for $\sigma = 1$.

Table \ref{table-compare} is a list of orbital parameters 
that are 
recovered by the Iwama+ approach and 
the present approach for $n_a = 100$, respectively.
In both cases, 
the difference between the true value of the semi-major axis and 
the mean of the recovered one decreases to less than a tenth.
This can be seen in Fig. \ref{comparing}, 
and is consistent with an order-of-magnitude estimation from
Eqs. (\ref{average-xx}) and (\ref{average-yy})
(see Appendix \ref{order-estimate}).
On the other hand, 
the dispersion of recovered parameters is not improved 
by the averaging operation
since the order of magnitude of the dispersion
depends on not $n_a$ but $\sigma/\sqrt{N}$.

In the case 1, the longitude of ascending node is well recovered 
with the accuracy of less than 10 \% of the true value,
while the other recovered parameters by the Iwama+ approach
are quite different from the true values.
By the averaging operation, all of mean recovered values approach 
the true values.
In the case 2, the recovered values except $\omega$ are improved.

Our numerical tests suggest that
$\omega$ and $\Omega$ are not always improved by the present approach.
However, this point is not important, 
since the change of the differences 
between the recovered values of $\omega$ and $\Omega$ 
and the true values of them 
are smaller than the dispersion of the recovered values.

In order to confirm the reliability of the above results, 
we calculate for $16 ( = 2^4)$ parameter sets as
$e_K = 0.1$ and $0.5$, $i = 30$ and $60$ [deg.], 
$\omega = 30$ and $60$ [deg.], and $\Omega = 30$ and $60$ [deg.].
One example as
$(e_K, i, \omega, \Omega) 
= (0.5, 60 \mathrm{~[deg.]}, 60 \mathrm{~[deg.]}, 60 \mathrm{~[deg.]})$ is
added into Table \ref{table-compare} for saving the space.
Fourier analyses recover the orbital period 
from numerically simulated data of the above two cases
with the accuracy of $\simeq$ 1 \% and 5 \%, respectively.

\subsection{Possible Application to Cyg X-1}

Let us consider a possible application to Cyg X-1, 
whose angular radius is $\sim 0.03$ milli-arcseconds (mas).
The required precision of the Small-JASMINE is $0.01$ mas,
so that $\sigma/\sqrt{N} = 0.01/0.03 \simeq 0.3$.
For Cyg X-1,
the Small-JASMINE is expected to measure the position of the star 
with the accuracy of 3 mas, which corresponds to $\sigma = 100$, 
for each imaging.
Hence, the position measurements of $N \simeq 10^5$ times are required 
as one data-set for $\sigma/\sqrt{N} \simeq 0.3$.

Since the Small-JASMINE is expected to measure 
for $3 - 4$ orbital periods of Cyg X-1, 
$\simeq 10^6$ observed points, 
which correspond to 10 data-sets, will be obtained. 
This means that $\sigma/\sqrt{N} \simeq 0.1$ exceeds 
the required precision of the Small-JASMINE.
However, every observed point has 
the systematic error of the Small-JASMINE as $\sim 0.01$ mas, 
so that the recovered parameters might also have the error of $\sim 0.01$ mas.

In this paper, we consider two cases as numerical tests
where we fix $\sigma/\sqrt{N}$ for each data-set.
Case 1: $N = 100$, $\sigma = 3$, $n_a = 1$. 
In this case, the present approach reduces to the Iwama+ one.
Case 2: $N = 1000$, $\sigma = 9.5$, $n_a = 10$.
See Table \ref{table-CygX-1} for a comparison of these cases
of one data-set.
In the case 2,
the present approach recovers 
the semi-major axis and the inclination better than the Iwama+ one.

On the other hand, 
the recovered eccentricity 
by the Iwama+ approach is close to the true value,
which is considered to be an accidental coincidence.
Numerical calculations of other parameter sets suggest
that the recovered eccentricities by the Iwama+ approach 
and the present one are close to $0.1$ and $0.25$, respectively, 
for any true eccentricity in the case 2.
Hence, the recovered eccentricities by the two approaches
may not be reliable when $\sigma/\sqrt{N} = 0.3$.
For the reliability of the recovered eccentricity, 
the position measurement 
with the accuracy of $\sigma/\sqrt{N} = 0.01$,
which corresponds to the case 1 in the Table \ref{table-compare},
is required.

The recovered values by the present approach 
are comparable to those by the Iwama+ approach
for the argument of periastron and the longitude of ascending node.
The recovered parameters by the present approach 
in the case 2
are comparable to the case 1.
These numerical results suggest that 
one can obtain the similar results 
for $\sigma = 3$ and $\sigma = 9.5$ by the averaging operation.
Hence, the present approach works well to reduce $\sigma$ effectively
for $n_m =100$.

Next, let us consider the same two cases for 10 data-sets.
Table \ref{table-CygX-1-10sets} shows 
the recovered values by the Iwama+ approach and the present approach
for 10 data-sets.
In both cases, 
each bin has 10 data-sets of $n_a$ observed points,
so that $10 \times n_a$ observed points are effectively averaged 
in the present approach.
On the other hand, 
every observed point is not averaged in the previous approach.

For the semi-major axis, 
the mean values of the recovered parameters 
by the previous approach of 10 data-sets 
in both cases are comparable to those of one data-set.
On the other hand, 
the recovered semi-major axis by the present approach of 10 data-sets
in both cases are much better than those of one data-set.

However, 
the recovered eccentricities by the both approaches
may not be reliable if $\sigma/\sqrt{N} = 0.3$
by the similar reason of one data-set.
For the reliability of the recovered eccentricity, 
the position measurement 
with the accuracy of $\sigma/\sqrt{N} = 0.01$,
which corresponds to the case 1 in the Table \ref{table-compare},
is required.
For the inclination,
the mean values of the recovered parameters by the present approach
are better than those by the previous approach.
The recovered values by the present approach 
are comparable to those by the previous approach
for the argument of periastron and the longitude of ascending node.
Note that
the dispersion of the recovered semi-major axis for 10 data-sets corresponds to
random errors of observations.
Therefore, the actual observational errors 
including the systematic errors of the Small-JASMINE 
will be comparable to the dispersion of Table \ref{table-CygX-1}.
These results suggest that 
the semi-major axis of Cyg X-1 is recovered 
with the accuracy comparable to or smaller than 
the true value of the semi-major axis
by the Small-JASMINE observations.
In order to search the best-fit parameter values,
recovered values by the present moment approach
would work well as trial values 
in the steepest descent method.

\subsection{Comparison with the inversion formula
by Asada, Akasaka, and Kasai (2004)}

As stated in section 1, 
Asada, Akasaka, and Kasai (2004) 
have found an exact solution for orbit determinations of astrometric binaries. 
The least square method is 
incorporated into the analytic solution
by Asada, Akasaka, and Kudoh (2007).
Their numerical calculations show that the analytic method recovers
orbital elements for small $\sigma$ cases, 
such as $\sigma = 0.001$ for $N=12$.

We also investigate the averaging operation
for the analytic solution 
\citep{AAK,AAK2,Asada2008}.
In the analytic solution, temporal information is 
fully considered through the law of constant-areal velocity.
In addition, one can use more points than $N/100$ 
for the averaging operations
because less than $100$ averaged points are required for 
the orbit determination differently from the moment approach.
Therefore, it seems that the accuracy of the parameter determination
by the analytic solution using the averaged points is 
better than that by the present approach.
However, numerical calculations suggest that it is not the case.
This is mainly because of two reasons: 
first, in the analytic solution, 
parameters of an apparent ellipse must be estimated
before the determination of orbital elements, 
and these parameters 
can not be recovered well
for an extremely low SN ratio.
Secondly, the eccentric anomaly $u$
that is needed for calculating areal velocities 
can not be recovered well.
Therefore, recovered values of the orbital elements 
can be complex numbers or quite different from the true values of them
(see Table \ref{AAK-2}),
where complex numbers would suggest hyperbolic orbits rather than elliptic one.
Hence, we do not make a further comparison 
between the present moment approach
and the inversion formula method.

\section{Conclusion}

This paper improved the Iwama+ approach for an extremely low SN ratio, 
where observational errors are comparable to or larger than
a binary size.
With avoiding a direct use of the time-consuming Kepler equation,
temporal information is taken into account 
to increase the accuracy of statistical moments.
As numerical tests, 100 realizations are done 
and the mean and the standard deviation are also evaluated.
For instance, 
the difference between the mean of the recovered values of the semi-major axis 
and the true value of that is decreased to less than a tenth 
in the case of $10000$ observed points.
Therefore, when one has a number of the observed points,
the present moment approach significantly improves the previous one
for the orbit determinations.
For Cyg X-1, 
the semi-major axis is expected to be recovered 
with the accuracy comparable to, or smaller than the true value
from astrometric observations alone.
Although the inversion formula by Asada, Akasaka, and Kasai is 
also discussed, 
numerical calculations show that 
the averaging operation
does not work well in the analytic method.
It is more convenient to start with 
values that are recovered by the present moment approach
and next use the steepest descent method 
for finally reaching the best-fit parameter values.
It is left as a future work.

We wish to thank the JASMINE science WG member 
for stimulating conversations. 
We would be grateful to Y. Sendouda and T. Yano
for useful discussions.
This work was supported 
in part (N.G.) by 
Ministry of Education, Culture, Sports, Science and Technology, 
Grant-in-Aid for Scientific Research (A), No. 23244034, 
and in part (K.Y.) by Japan Society for the Promotion of Science, 
Grant-in-Aid for JSPS Fellows, No. 24108.

\appendix

\section{Estimation of the recovered semi-major axis}
\label{order-estimate}

Let us estimate the difference between the recovered value and true one
for the semi-major axis.
Using Eqs. (\ref{alpha}) - (\ref{delta}), (\ref{Mxx}), and (\ref{Myy}), 
Eqs. (\ref{error-xx}) and (\ref{average-xx}) are rewritten as
\begin{eqnarray}
E \left( f(e_K^{(O)}, i^{(O)}, 
\omega^{(O)}, \Omega^{(O)} )
(a_K^{(O)})^2  \right)
&=& 
f (e_K^{(T)}, i^{(T)}, \omega^{(T)}, \Omega^{(T)} ) (a_K^{(T)})^2
+ \frac{N-1}{N} \sigma^2 ,
\label{ex-faK1}
\\
E \left( f(\bar{e}_K^{(O)}, \bar{i}^{(O)}, 
\bar{\omega}^{(O)}, \bar{\Omega}^{(O)} )
(\bar{a}_K^{(O)})^2  \right)
&=& 
f (e_K^{(T)}, i^{(T)}, \omega^{(T)}, \Omega^{(T)} ) (a_K^{(T)})^2
+ \frac{N-1}{N} \frac{\sigma^2}{n_a} ,
\label{ex-faK2}
\end{eqnarray}
where 
\begin{eqnarray}
f (e_K, i, \omega, \Omega) &\equiv&
\frac12 [ (\cos \omega \cos \Omega - \sin \omega \sin \Omega \cos i)^2
+ (1 - e_K^2) (\sin \omega \cos \Omega + \cos \omega \sin \Omega \cos i)^2 ]
\nonumber\\
&&
- \frac14 e_K^2 (\cos \omega \cos \Omega - \sin \omega \sin \Omega \cos i)^2.
\end{eqnarray}
In the order-of-magnitude estimation, 
one can assume that $a_K$ and $f (e_K, i, \omega, \Omega)$
are independent of each other.
Hence, 
\begin{eqnarray}
E( f(e_K, i, \omega, \Omega) a_K^2 ) \simeq 
E( f(e_K, i, \omega, \Omega)) E( a_K^2 ). 
\end{eqnarray} 
In addition, 
because $e_K$ and the trigonometric functions are from $0$ to $1$,
one finds $f (e_K, i, \omega, \Omega) = \mathrm{O}(1)$.
Therefore, 
the expected values of the recovered semi-major axis 
by the Iwama+ approach and by the present one
are expressed approximately as
\begin{eqnarray}
E \left( a_K^{(O)} \right)
&\sim&  \sqrt{\left(a_K^{(T)} \right)^2 + \sigma^2}
\geq a_K^{(T)} + \sigma,
\label{ex-aK1}
\\
E \left( \bar{a}_K^{(O)} \right)
&\sim& \sqrt{\left(a_K^{(T)} \right)^2 + \frac{\sigma^2}{n_a} }
\geq a_K^{(T)} + \frac{\sigma}{\sqrt{n_a}} ,
\label{ex-aK2}
\end{eqnarray}
respectively.
Eqs. (\ref{ex-aK1}) and (\ref{ex-aK2}) suggest that
the difference between the true value of the semi-major axis 
and the mean value of the recovered one 
decreases to nearly $1/\sqrt{n_a}$ by the averaging operation.
If $n_a \geq 100$, this difference decreases to nearly a tenth.


\clearpage

\clearpage

\clearpage

\begin{table}[h]
\caption{Reconstructing two parameter sets of numerical simulations 
for two different cases by the Iwama+ and the present approaches.
In the table, the row $\sigma = 0$ indicates true orbital parameters, 
whereas the rows $\sigma = 1$ and $5$ 
provide the recovered values for adding Gaussian errors 
($1$ or $5$ in the units of the true semi-major axis, respectively). 
The total number of the observed points $N = 10000$ and $n_a = 100$.
For each parameter set, 100 realizations are done and 
the mean and the standard deviation are also evaluated. 
For the semi-major axis, 
the difference between the mean of the recovered values 
and the true value decreases to less than a tenth.}
\begin{center}
\begin{tabular}{l|l|lllll}
$\sigma$ & Approach & $a_K$ & $e_K$ & $i$ [deg.] & 
$\omega$ [deg.] & $\Omega$ [deg.] \\ 
\hline
\hline
$0$ & & $1.0$ & $0.1$ & $30$ & $30$ & $30$ \\
\hline
$1$ & Iwama+ & $1.73 \pm 0.01$ & $0.030 \pm 0.017$ & 
$17.2 \pm 1.9$ & $36.6 \pm 24.7$ & $30.9 \pm 6.1$ \\
 & Present & $1.01 \pm 0.01$ & $0.103 \pm 0.032$ & 
$30.0 \pm 2.2$ & $29.6 \pm 17.8$ & $30.2 \pm 4.5$ \\
\hline
$5$ & Iwama+ & $7.18 \pm 0.04$ & $0.027 \pm 0.014$ & 
$9.1 \pm 2.3$ & $44.5 \pm 26.7$ & $39.9 \pm 25.9$ \\
 & Present & $1.27 \pm 0.07$ & $0.203 \pm 0.115$ & 
$28.7 \pm 7.0$ & $49.7 \pm 27.3$ & $32.3 \pm 20.4$ \\
\hline
\hline
$\sigma$ & Approach & $a_K$ & $e_K$ & $i$ [deg.] & 
$\omega$ [deg.] & $\Omega$ [deg.] \\ 
\hline
\hline
$0$ & & $1.0$ & $0.5$ & $60$ & $60$ & $60$ \\
\hline
$1$ & Iwama+ & $1.67 \pm 0.01$ & $0.054 \pm 0.021$ & 
$27.1 \pm 1.1$ & $42.4 \pm 23.6$ & $62.5 \pm 2.5$ \\
 & Present & $1.00 \pm 0.03$ & $0.469 \pm 0.049$ & 
$58.4 \pm 1.2$ & $58.8 \pm 5.5$ & $59.9 \pm 1.4$ \\
\hline
$5$ & Iwama+ & $7.16 \pm 0.04$ & $0.029 \pm 0.015$ & 
$9.3 \pm 2.8$ & $43.9 \pm 26.5$ & $54.4 \pm 24.3$ \\
 & Present & $1.19 \pm 0.09$ & $0.282 \pm 0.140$ & 
$41.7 \pm 7.4$ & $53.6 \pm 25.7$ & $60.4 \pm 8.8$ \\
\hline
\hline
\end{tabular}
\end{center}
\label{table-compare}
\end{table}

\clearpage

\begin{table}[h]
\caption{Reconstructing the parameters of numerical simulations 
for two different cases as ($N=100$, $\sigma = 3$, $n_a=1$)
and ($N=1000$, $\sigma = 9.5$, $n_a=10$) of one data-set. 
In the table, the row $\sigma = 0$ indicates the true orbital parameters, 
whereas the rows $\sigma = 3$ and $9.5$  
provide the recovered values for adding Gaussian errors 
($3$ or $9.5$ in the units of the true semi-major axis, respectively). 
For each parameter set, 100 realizations are done and 
the mean and the standard deviation are also evaluated. } 
\begin{center}
\begin{tabular}{l|l|l|lllll}
$\sigma$ & $N$ & Approach & 
$a_k$ & $e_k$ & $i$ [deg.] & $\omega$ [deg.] & $\Omega$ [deg.] \\ 
\hline
\hline
$0$ & $100$ & & $1.0$ & $0.1$ & $30$ & $30$ & $30$ \\
\hline
$3$ & $100$ & Iwama+ & $4.76 \pm 0.41$ & $0.285 \pm 0.133$ & 
$27.2 \pm 8.3$ & $51.9 \pm 25.0$ & $41.4 \pm 23.5$ \\
\hline
$9.5$ & $1000$ & Iwama+ & $13.77 \pm 0.27$ & $0.086 \pm 0.045$ & 
$15.7 \pm 4.3$ & $48.0 \pm 25.1$ & $41.4 \pm 27.8$ \\
 & & Present & $4.78 \pm 0.39$ & $0.254 \pm 0.142$ & 
$28.1 \pm 7.3$ & $55.3 \pm 24.5$ & $38.4 \pm 24.6$ \\
\hline
\hline
\end{tabular}
\end{center}
\label{table-CygX-1}
\end{table}

\begin{table}[h]
\caption{Reconstructing the parameters of numerical simulations 
for two different cases as ($N=100$, $\sigma = 3$, $n_a=1$)
and ($N=1000$, $\sigma = 9.5$, $n_a=10$) of 10 data-sets. 
In the table, the row $\sigma = 0$ indicates the true orbital parameters, 
whereas the rows $\sigma = 3$ and $9.5$  
provide the recovered values for adding Gaussian errors 
($3$ or $9.5$ in the units of the true semi-major axis, respectively). 
For each parameter set, 100 realizations are done and 
the mean and the standard deviation are also evaluated. } 
\begin{center}
\begin{tabular}{l|l|l|lllll}
$\sigma$ & $N$ & Approach & 
$a_k$ & $e_k$ & $i$ [deg.] & $\omega$ [deg.] & $\Omega$ [deg.] \\ 
\hline
\hline
$0$ & $100$ & & $1.0$ & $0.1$ & $30$ & $30$ & $30$ \\
\hline
$3$ & $100$ & Iwama+ & $4.43 \pm 0.08$ & $0.090 \pm 0.046$ & 
$15.0 \pm 3.8$ & $46.3 \pm 23.9$ & $45.4 \pm 25.8$ \\
 & & Present & $1.77 \pm 0.11$ & $0.255 \pm 0.125$ & 
$26.6 \pm 8.0$ & $53.1 \pm 27.1$ & $38.7 \pm 24.7$ \\
\hline
$9.5$ & $1000$ & Iwama+ & $13.56 \pm 0.08$ & $0.031 \pm 0.016$ & 
$9.1 \pm 2.5$ & $46.9 \pm 27.0$ & $43.2 \pm 25.7$ \\
 & & Present & $1.79 \pm 0.13$ & $0.236 \pm 0.137$ & 
$28.0 \pm 7.1$ & $52.0 \pm 23.0$ & $35.1 \pm 23.1$ \\
\hline
\hline
\end{tabular}
\end{center}
\label{table-CygX-1-10sets}
\end{table}

\begin{table}[h]
\caption{
Reconstructing the parameters 
for the analytic solution using the averaged points.
In the table, the $\sigma = 0$ row indicates true orbital parameters,
whereas the $\sigma = 1$ row provides 
the mean of the recovered parameters by the inversion formula
for $N = 100$ observed points.
$10$ points are used by the averaging operation
for each averaged point, 
and 100 realization are done.
The orbital elements are 
$a_k=1.0$, $e_k = 0.1$, $i = 30$[deg.], $\omega = 30$[deg.], and 
$\Omega = 30$[deg.], respectively.} 
\begin{center}
\begin{tabular}{l|l|lllll}
$\sigma$ & $N$ & 
$a_k$ & $e_k$ & $i$ [deg.] & $\omega$ [deg.] & $\Omega$ [deg.] \\ 
\hline
\hline
$0$ & $100$ &
$1.0$ & $0.1$ & $30$ & $30$ & $30$ \\
\hline
$1$ & $100$ & 
$3.06 $&
$13.065 $ &
$78.7 + 17.0 i $&
$7.5 + 0.8 i $&
$51.3 $
\\
\hline
\hline
\end{tabular}
\end{center}
\label{AAK-2}
\end{table}

\end{document}